\documentclass{aa}
\usepackage{psfig,latexsym,longtable,lscape}

\newcommand{\D}{$^\circ$}
\newcommand{\ESO }{\mbox {ESO\,295-IG022}}
\newcommand{\ESOs }{\mbox {ESO\,295-IG022-S}}
\newcommand{\ESOn }{\mbox {ESO\,295-IG022-N}}
\newcommand{\CL }{\mbox {Cl\,0053-37}}
\newcommand{\AT }{\mbox {ATCA\,J0056.0-3723}}
\newcommand{\AB }{\mbox {Abell\,S0102}}

\newcommand{\etal}{et al.}

\newcommand{\eg}{e.g.}
\newcommand{\ie}{i.e.}

\def\arcm{\hbox{$^\prime$}}

\def\p0{\phantom{0}}

\begin{document}

%
   \title{Radio jets and diffuse X-ray emission around the peculiar galaxy pair 
	  ESO\,295-IG022}

   \author{A.M.\,Read\inst{1} \and M.D.\,Filipovi\'c\inst{1,2,3} 
           \and W.\,Pietsch\inst{1} \and P.A.\,Jones\inst{2}
          }

   \offprints{A.M.\,Read: aread@mpe.mpg.de}

   \institute{
   Max-Planck-Institut f\"ur extraterrestrische Physik, 
    Postfach 1312, D-85741 Garching, Germany
   \and
   University of Western Sydney Nepean, 
    P.O. Box 10, Kingswood NSW 2747, Australia
   \and
   Australia Telescope National Facility, 
    CSIRO, P.O. Box 76, Epping NSW 2121, Australia
   \\
   email: aread@mpe.mpg.de ; m.filipovic@uws.edu.au ; wnp@mpe.mpg.de ; p.jones@uws.edu.au  \\
              } 
   \date{Received -- ; accepted --}

	\authorrunning {A.M.\,Read \etal}
	\titlerunning {Radio jets and diffuse X-ray emission around the 
          peculiar galaxy pair ESO\,295-IG022}

\abstract{
We report Australia Telescope Compact Array (ATCA) radio-continuum and ROSAT
PSPC X-ray observations of the region surrounding the peculiar galaxy pair
\ESO, lying at the centre of the poor cluster \AB. We identify a radio galaxy
coincident with the galaxy pair, with bipolar, bent radio jets appearing
centred on the southern galaxy, and extending both to the south, for about
95\arcsec ($\approx100$\,kpc at the distance of \ESO), and to the north, for
$\approx80$\arcsec, encompassing the northern galaxy. We discuss also the idea
of an additional single jet structure from the northern galaxy contributing to
the emission. We estimate lower limit jet velocities of at least 1000\,km
s$^{-1}$, and a relative proper velocity for the southern galaxy through the
cluster of $\sim200$\,km s$^{-1}$.  Diffuse ($kT=2.2$\,keV) X-ray emission
consistent with group or poor cluster emission is seen centred on the southern
galaxy surrounding the galaxy pair and the associated radio jets. Structure
within the radio jets and the X-ray emission is very suggestive of there being
some channeling of the radio emission with the surrounding intragroup medium.
Another bipolar jet radio galaxy, discovered close by, is likely to be a
background object, the optical counterpart having a magnitude m$>$22.
}
\maketitle


\section{Introduction}

Radio galaxies in clusters and groups are an important probe of the
intracluster medium (ICM), as the jets can be confined and bent by the dense
medium, and the non-thermal synchrotron radio emission provides complementary
data to the (mainly thermal) X-ray emission.

The galaxy cluster \AB\ (Abell \etal\ 1989) (also known as EDCC~494; Lumsden
\etal\ 1992), at a redshift of z=0.054824 (Bica \etal\ 1991), is a poor
cluster from Abell's supplementary catalogue. The \ESO\ galaxy pair (Bica
\etal\ 1991), classified as a merging galaxy system (Lumsden \etal\ 1992) and
members of the APM Bright Galaxy Catalogue (Loveday 1996), lie at the
cluster centre. Cappi \etal\ (1998) showed that the southeastern part of \AB\
is in fact a separate cluster (\CL) at a more distant redshift (z=0.165), and
is therefore not physically connected with the rest of \AB. Lumsden \etal\
(1992) estimated that \AB\ has some 15+ members within a radius of
$\sim$20$\arcmin$, and recently, Ratcliffe \etal\ (1998), in their
'Durham/UKST Galaxy Redshift Survey', gave optical details of some 78 \AB\
cluster members.

Although the cluster was covered by the NVSS radio-continuum survey, the low
resolution (40$\arcsec$) meant that the radio structures were not resolved
clearly. This galaxy cluster however, lies some 18$\arcmin$ northeast of the
nearby Sculptor group galaxy NGC~300, and radio observations of NGC~300 with
the Australia Telescope Compact Array (ATCA) and X-ray observations with the
ROSAT PSPC, also covered the area of this cluster, lying within the primary
beam of the ATCA $\lambda$=20~cm wavelength and within the ROSAT PSPC field.
Here, we present these new ATCA and ROSAT PSPC results. In Sect.\,2, we 
describe the radio, X-ray and optical observations and data analysis. The 
results of this analysis are presented in Sect\,3, and discussed in Sect\,4. 
Finally, in Sect\,5, we present our conclusions.

\section{Observations and data analysis}

  \subsection{Radio-continuum data}

The \AB\ region was observed as part of the NGC~300 ATCA observations with a
baseline of 6~km at 1384~MHz ($\lambda$=20~cm) and 2496~MHz ($\lambda$=13~cm),
with corresponding angular resolutions of 6$\arcsec$ and 4$\arcsec$. More
information regarding these observations can be found in Filipovi\'c et al.
(in preparation). These observations were quite far
down the primary beam pattern of the ATCA, and so some primary beam correction
has been applied using the standard
techniques in the 
\mbox{MIRIAD} software package (Sault \& Killeen 1999) 
with the parameters of the ATCA primary beam. 

Additional archival 4800\,MHz ($\lambda=6$\,cm) ATCA data, centred on \ESO,
was also used, though only of $\sim$20 minutes duration and of fairly low
sensitivity. This data was used only to determine the main feature's 
core flux and position.

1384~MHz ($\lambda$=20~cm) primary beam corrected 
radio contours of these data are shown in Fig.1,
overlaid on a Digital Sky Survey (red) image. Two bipolar jet radio sources
are seen, one (J0055.7-3724: 
core at $\alpha$(J2000)=00$^{\rm h}$55$^{\rm m}$46.58$^{\rm s}$,
$\delta$(J2000)=-37$^{\circ}$24\arcm27.7\arcsec) coincident with the southern
galaxy (\ESOs), the other (J0056.0-3723: 
core at $\alpha$(J2000)=00$^{\rm h}$56$^{\rm
m}$00.21$^{\rm s}$, $\delta$(J2000)=-37$^{\circ}$23\arcm47.6\arcsec) having a
very faint optical counterpart (positions are from the 1384~MHz observations).
The frequency-dependent integrated flux densities ($S$) and, where applicable,
lengths ($L$) of J0055.7-3724's core and two jet features, and of the entirety
of J0056.0-3723 are given in Table~\ref{tab-rsrc}. As the 13 and 6~cm data are
of lower signal-to-noise ratio than the 20~cm data, these corresponding fluxes
for the low surface brightness northern and southern jets (and of
J0056.0-3723) have been omitted. The 13~cm J0055.7-3724 core flux value is
also rather uncertain due to the beam correction applied.

\begin{figure*}[t]
    \psfig{figure=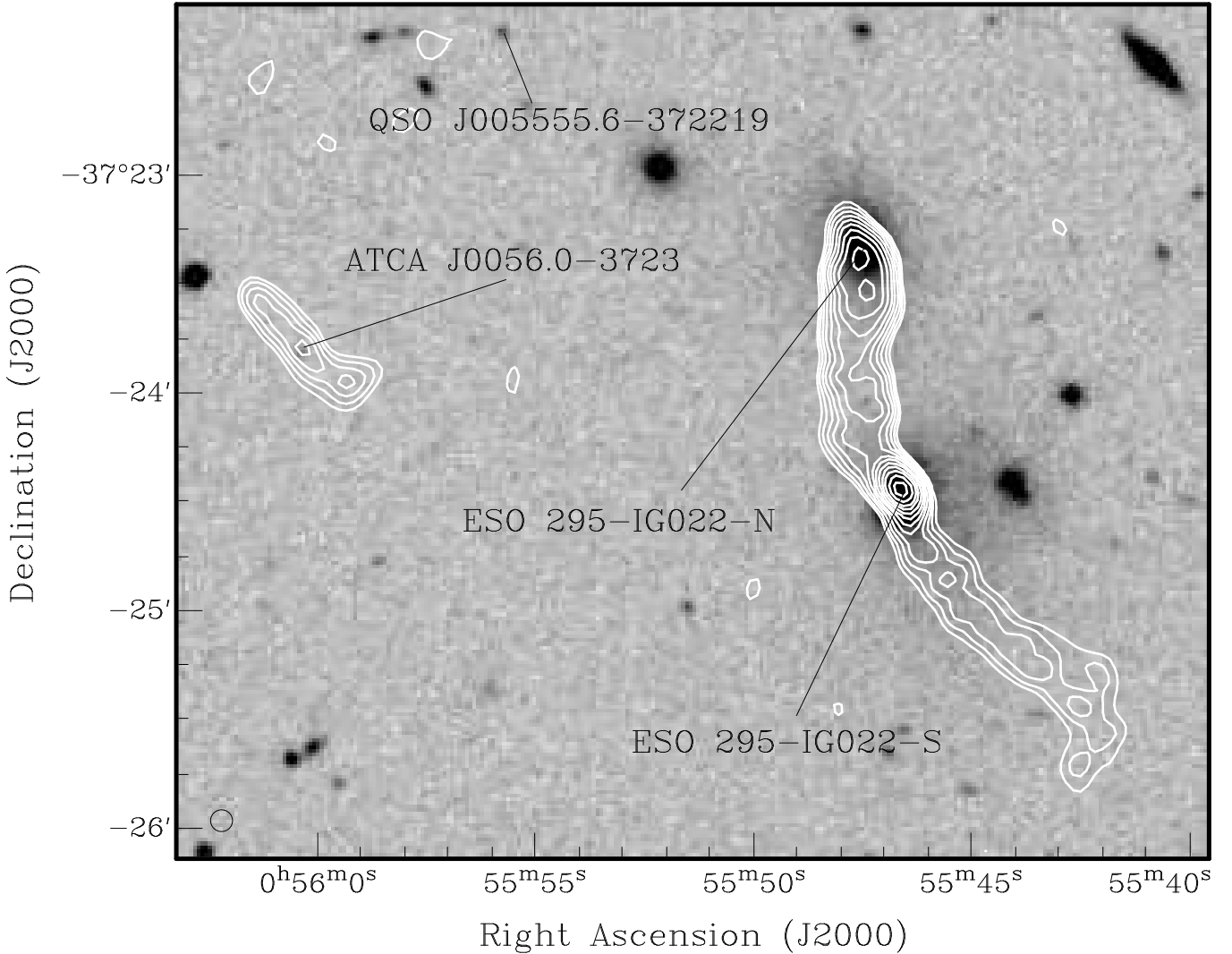,width=18cm,clip=}
{\caption {Digital Sky Survey (red) image of the
Abell~S0102 cluster central region overlaid with 
primary beam corrected  1384\,MHz ATCA radio-continuum contours. 
The synthesized beam of the ATCA observation is
6$\arcsec$ $\times$ 6$\arcsec$ (see circle, lower left corner) with r.m.s. 
noise (1$\sigma$) of 0.06~mJy. Contours increase by factors of $\sqrt{2}$ 
from 0.5 mJy/beam
}}
\end{figure*}
%

\begin{table}[t]
\caption[]{Radio-continuum integrated flux
densities ($S$) and lengths ($L$) of the two radio
galaxies within Abell\,S0102.}
 \begin{flushleft}
  \begin{tabular}{lcccc}
\hline
\noalign{\smallskip}
ATCA         & L     & 
$\rm S_{\rm 20\,cm}$ & $\rm S_{\rm 13\,cm}$ & $\rm S_{\rm 6\,cm}$    \\
Radio Name   & ($\arcsec$) &
 (mJy) & (mJy) & (mJy)  \\
\hline
\noalign{\smallskip}
J0055.7-3724 (core)  & -- & \p047.0 & 63.9 & 32.6 \\
J0055.7-3724 (N jet) & 80 & 103.0   & --   & --   \\
J0055.7-3724 (S jet) & 95 & \p036.7 & --   & --   \\
J0056.0-3723 (all)   & 45 & \p017.6 & --   & --   \\
\hline
  \end{tabular}
 \end{flushleft}
\label{tab-rsrc}
\end{table}

The J0055.7-3724 20~cm integrated flux density was determined as the sum of
the flux density in a box around \ESOs\ and the two jet features. The flux
density of the central radio core coincident with \ESOs\ was determined using
the two-dimensional elliptical Gaussian fitting algorithm within the
\mbox{MIRIAD} software package. Similarly, flux
densities of the \ESOs\ core at 13 and 6~cm were determined once the images
were smoothed to the same resolution as the 20~cm data, \ie\ to 6$\arcsec$.
Our derived integrated flux at 20\,cm
for the ESO 295-IG022 source agrees well with the flux from the VLA NVSS.

  \subsection{\mbox{X-ray} data}

The field surrounding NGC~300 was observed with the ROSAT PSPC (see
Tr\"{u}mper 1982) twice; for 9324 seconds in November 1991, and for 36693
seconds in May 1992. Each dataset was cleaned of both very high and very low
accepted event rates and master veto rates. Source detection and position
determination were performed over the full field of view with the EXSAS local
detect, map detect, and maximum likelihood algorithms (Zimmermann et al.
1994). The two eventsets were then shifted with respect to the prominent
bright star in the field, HD5403, correcting for the proper motion of the star
(Perryman \etal\ 1997) at the epoch of the ROSAT observations. The two cleaned
and position-corrected datasets were then merged together, and the source
detection procedures were re-ran yielding 79 sources. Though many sources were
found within NGC~300 (detailed in a forthcoming paper, Read et al., in preparation),
much interesting \mbox{X-ray} structure was seen surrounding the interacting
galaxy pair \mbox{ESO\,295-IG022}, and we concentrate here solely on this
emission.

As, unfortunately, this area of the sky was partially obscured by the ROSAT
PSPC entrance window support structure, a careful exposure correction had to be
performed. Exposure maps were calculated in seven different standard ROSAT
energy bands. These were combined with cleaned images in the same seven bands,
and then smoothed with Gaussians on energy-dependent spatial scales (using a
smoothing $\sigma$ of from 66$\arcsec$ [low energy] to 33$\arcsec$ [hard
energy]). The resultant cleaned, exposure-corrected and smoothed images were
combined into broad \mbox{(0.1$-$2.4\,keV)}, soft \mbox{(0.1$-$0.4\,keV)},
hard \mbox{(0.5$-$2.0\,keV)}, hard~1 \mbox{(0.5$-$0.9\,keV)} and hard~2
\mbox{(0.9$-$2.0\,keV)} band images. Fig.\,2 shows the resultant hard-band
\mbox{(0.5$-$2.0\,keV)} image as grey scale, overlaid with the 1384\,MHz
J0055.7-3724 radio contours of Fig\,1. The X-ray emission is seen to surround
the J0055.7-3724 radio feature and, though 
no obvious central peak to the emission is 
evident, bright clumps are seen, notably in between the two \ESO\ galaxies, 
and both to the northwest of the southern jet and to the 
east of the southern jet `foot'. 
Note that the three ROSAT HRI
observations of NGC~300 were not used, as the sensitivity at the \AB\ position
(at the very edge of the FOV) was severely reduced.

\begin{figure}
    \psfig{figure=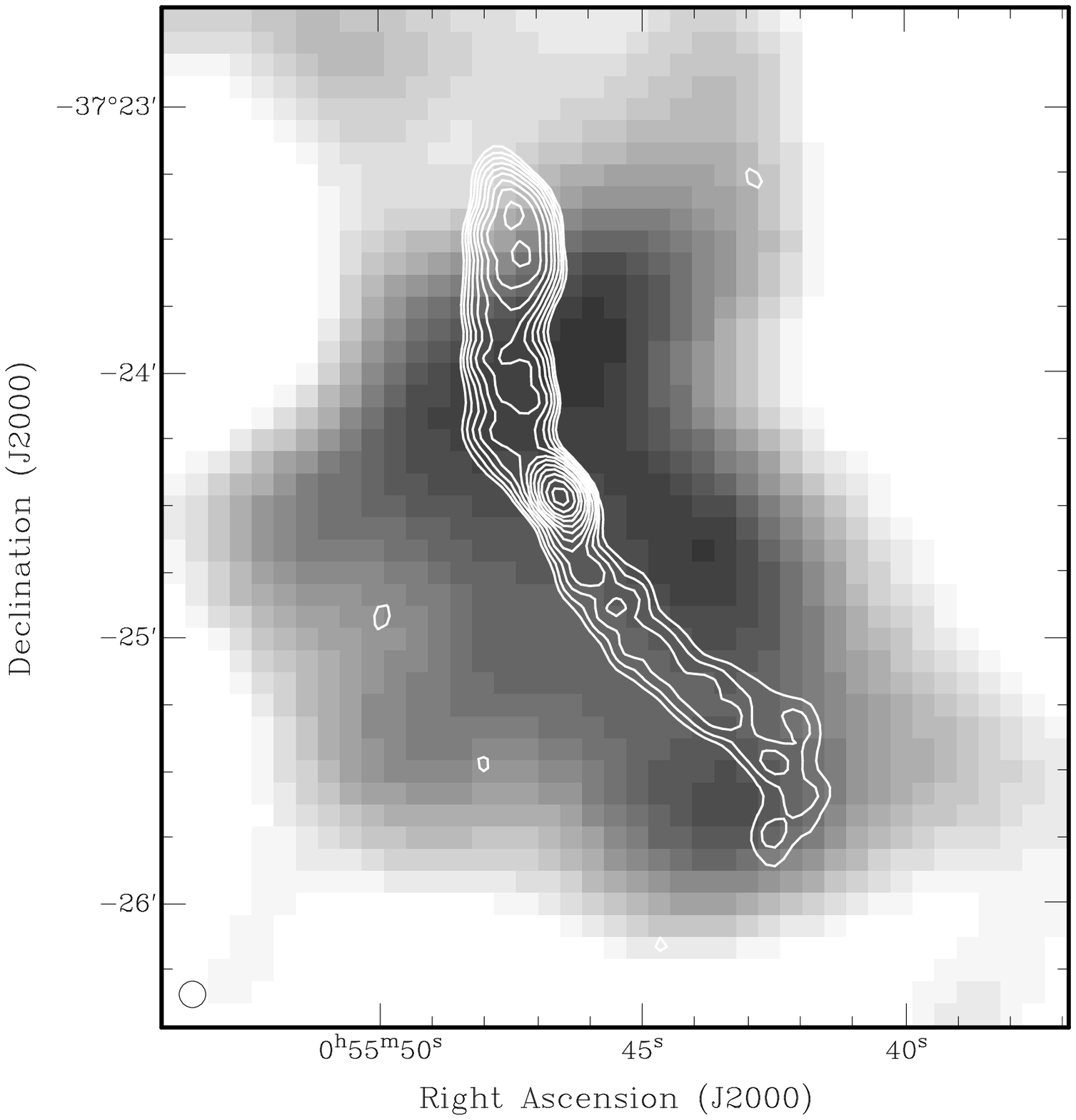,width=8.8cm,clip=}
\caption{
ESO 295-IG022 in detail: ROSAT hard X-ray band (0.5-2.0 keV) 
PSPC image overlaid with ATCA primary beam corrected 1384 MHz radio-continuum
contours (contour levels same as in Fig.~1). 
}
\end{figure}

 \subsection{Optical data}

In the Digital Sky Survey 2 (DSS2) optical red image of the field surrounding
\ESO\ (Fig.~1), we can see that the southern galaxy is {\em itself}
interacting, optical filaments being visible to the west, enveloping a nearby
companion. This can be seen far more clearly in Fig.\,3, where in addition,
two nuclei are seen within \ESOs\ itself. Looking into the 
lower-quality DSS1 data, taken 14 years earlier, we see that the situation 
appears identical, indicating that we have a true double nucleus system, 
and not some transient bright source such as a supernova. 
Surrounding \ESO, we see several other fainter optical galaxies, 
probable members of \AB.

\begin{figure}
    \psfig{figure=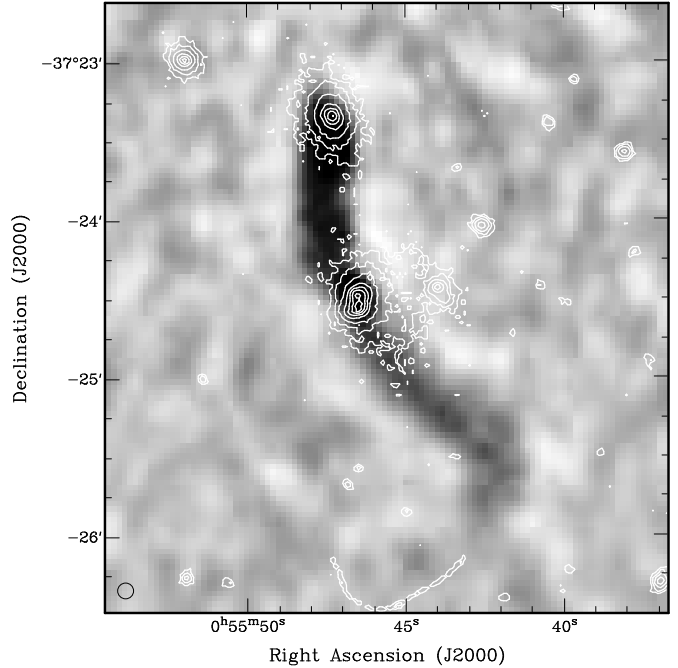,width=8.8cm,clip=}
\caption{
ESO 295-IG022 in detail: The Fig.~2 (and 1) primary beam corrected ATCA
radio-continuum image (this time as grey scale) of ESO~295-IG022, with optical
DSS2 (red) contours. The inner contour indicates the merging nuclei of
ESO~295-IG022-S
}
\end{figure}

\section {Discussion}

\subsection{ATCA\,J0055.7-3724}

In comparing the radio continuum, ROSAT PSPC and optical DSS2 images (Figs.\,1
and 2), we see some striking features in and around the central \AB\ cluster
galaxy pair, \ESO. The \ESOs\ core has a flat or inverted spectrum, but the
comparison of the data at the three frequencies (Table~1) indicates that the
core is probably variable, so the spectral index is difficult to determine.

The galaxy \ESOs\ is complex and disturbed with two nuclei, $\approx6$\arcsec\
(\ie\ 6.5\,kpc) apart, and a nearby interacting system around 30$\arcsec$ to
the west. The radio core is seen to lie within $1$\arcsec\ of the
northernmost \ESOs\ nucleus. The bright northernmost radio knot
($\alpha$(J2000)=00$^h$55$^m$47.53$^s$,
$\delta$(J2000)=$-$37\D23\arcmin22.5\arcsec) lies some $3-4$\arcsec\ from
\ESOn\, and so, given the positional errors of both the optical and radio data
(both $\approx$1\arcsec), it seems sensible to interpret the radio source as a
bipolar twin-jet with a core coincident with \ESOs. However, within the jets,
several knots of emission are seen, and the northern jet contains almost 3
times as much flux as the southern (see Table\,1), hence an alternative
interpretation, in that some of the northern radio emission originates 
as a single jet structure from the
northern galaxy, is plausible and will need
to be investigated in dedicated, new 6 and 3~cm ATCA observations.
Note that we do not have any reliable spectral information from the northern 
end of the radio jet, as the feature, though seen in the 20~cm data, is barely 
visible in the 13 and 6~cm data. 

The fact that the main `bend' in the radio emission appears to lie some
10\arcsec$-$15\arcsec\ northeast of the main core, adds some credence to this
second `two galaxies' hypothesis (\ie\ assuming jets from both \ESO\ 
galaxies), as one would expect in the `single galaxy'
model (\ie\ just a jet solely from \ESOs), 
to a first approximation, the bend to occur at the core, as the jet is
expelled from both sides and is swept back. However, it is well known (\eg\ 
Owen \& Ledlow 1997)
that bends occur in the radio features of cluster radio galaxies in a lot of
places, not just at the core, and a 
very similar offset bend to the present case
is found in the famous 3C\,465 radio galaxy at the centre of the the cluster
Abell~2634 (Sakelliou \& Merrifield 1999). Furthermore, it is known that there
are regions of brightening within radio jets associated with shocks and
re-acceleration (Hardee 1996), and this could well be the cause of the feature. 
Interestingly, the X-ray data do show a bright clump of denser emission close 
to the position of the bend, and so the bend may have been caused by 
the jet shocking into this denser material. 
All in all though, there are some arguments for the radio feature not being due 
solely to a twin-jet structure from \ESOs, but for there being some
component of a single jet structure from \ESOn\ as well. Further investigation
is necessary, and is in progress.

Though we have no real idea as to the orientation of the jets with our line 
of sight, if we assume them to lie essentially perpendicular to it, then some 
interesting inferences can be made. Assuming an $H_{0}$ of 75\,km s$^{-1}$
Mpc$^{-1}$, the \ESO\ redshift corresponds to a distance of 
$\approx220$\,Mpc, leading to jet lengths (or lower limits rather, given 
the uncertainty in the orientation) of about 85 (North) and 100 (South) kpc. 

In order to convert these distances into velocities, some assumed 
timescale, over which the jets are produced, is necessary. 
While we agree that there are certainly a few different 
methods of estimating these sorts of timescales, given the fact that both 
the radio data and the X-ray data are not of the highest possible quality (both 
having been obtained serendipitously, and suffering from effects to do with 
\ESO\ not being the prime target of the observations), we here use a simple 
method of estimation. With our upcoming observations, more complex methods 
for estimating the jet production timescale may be possible 
(note of course that the velocities calculated  
here are easily scaled with any other timescale assumed). 

Given that the two \ESOs\ nuclei are so close together, it seems sensible to
assume that \ESOs\ is at a similar interaction stage as mergers such as
Arp~220 and NGC~2623 (see Read \& Ponman 1998). This interaction stage is
thought both observationally and theoretically (\eg\ Mihos \& Hernquist 1996)
to give rise to very violent and efficient compressing and fueling of gas
towards the galaxies' centres. This vigorous gas fueling process towards the
galaxy centres is thought to last at most $10^{8}$\,yr, and assuming, as is
commonly believed, the radio structures observed to be due to compact AGN
activity (\eg\ Urry \& Padovani 1995; Laing 1993), here fed with gas via this
fueling process, then this interaction and gas-fueling timescale gives rise to
jet proper motions of 830$-$990\,km s$^{-1}$. The true jet speeds may be far
higher due to inclination effects. Also the jet timescale may be somewhat
shorter (note that the \ESOs\ nuclei have not yet completely merged). Note
also that the velocity of the end of the jet (which we are here estimating),
pushing through the medium, is probably slower than the flow of material
within the jet. As we do not have reliable radio spectral index gradients, it
is difficult to determine exact particle velocities. For the synchrotron
ageing model (Myers \& Spangler 1985), velocities of $\sim$1000\,km
s$^{-1}$ over timescales of $10^{8}$\,yr are entirely plausible. It is likely
however that re-acceleration does occur, and so synchrotron ageing may not be
relevant. Note also that synchrotron lifetimes may not be that reliable (Eilek
1996).

It can be further seen that 
a circle of radius 140\,kpc passes nicely through the radio core
and the bright southernmost and northernmost knots, following well the curve
of the radio jets. Standard radio jet creation models (\eg\ Blandford 1990)
predict straight jets. However, as is seen here, they are often bent 
(see also Owen \& Ledlow 1997), and this
is usually attributed to the motion of the galaxy through the ICM, resulting in
significant ram pressure stripping acting on the jets (Begelman \etal\ 1984).
Assuming \ESOs\ to move east-southeasterly, one can calculate a proper motion
(upper limit) of \ESOs\ of 190\,km s$^{-1}$, similar to the velocity
dispersion of poor clusters and groups. Note that this proper motion is often
attributable to the gravitational influence of a companion galaxy. Here, the
position of the westerly companion and the interaction-induced optical
filaments (Fig.\,3) are nicely aligned with the proper motion of \ESOs.

The ROSAT PSPC hard-band image (Fig.~2) shows much emission enveloping the
\ESO\ galaxy pair. To analyse the cluster/group X-ray emission,
a 400$\arcsec$ diameter region centred 
5\arcsec\ north of the southern galaxy was selected. Comparing the broad-band
(unsmoothed) exposure-corrected image with source-subtracted background
regions to the northeast and northwest, a broad-band background-subtracted
countrate of \mbox{0.09~cts/s} was estimated. A source spectrum, extracted
from the same region was fitted with standard spectral models (power law,
thermal bremsstrahlung, blackbody and Raymond \& Smith hot plasma models). A
best fit (reduced \mbox{$\chi^{2}$ = 0.69}) was obtained with a
low-temperature (\mbox{kT=2.2\,keV}) Raymond \& Smith hot plasma model, with
the metallicity frozen at 0.2 solar. The fitted absorbing column was found to
be entirely consistent with the Galactic value in this direction
(\mbox{2.97$\times$10$^{20}$\,atoms cm$^{-2}$}; Dickey \& Lockman 1990),
indicating no intrinsic absorption is present. Combining the results of the
spectral fitting with the calculated exposure-corrected countrate and the
redshift of \AB, one arrives at an \mbox {X-ray} luminosity
(\mbox{0.1$-$2.4\,keV}) of \mbox{$8.6\times10^{42}$\,erg s$^{-1}$} for this
extended emission. This luminosity, combined with the low temperature and
metallicity and the lack of intrinsic absorption, fits very 
well with the idea that
the emission is due to galaxy group/small cluster (as opposed to large
cluster) emission, the emission being due to hot gas lying in the potential
well of the group, of which \mbox{ESO\,295-IG022-N/S} are the prominent
members. Note that this idea sits very nicely 
with the fact that the proper motion of 
\ESOs\ calculated earlier is also more similar to that of groups than of large 
clusters. 

The comparison between the radio and X-ray emission (Fig.\,2) is very
suggestive of the southern jet being confined in the \mbox{X-ray} gas. There
is an indication of an X-ray radio anticoincidence around the southern jet,
particularly to the northwest near the core and to the east near the southern
tip. The jets may be collimated by the surrounding X-ray `cocoon', which
maintains the pressure of the ICM in balance with that of the gas within the
jet. This would be very suggestive of channeling effects taking place, whereby
the violent radio-continuum jets are able to punch holes and displace the
X-ray emitting cluster gas, as is seen in the NGC~1275 jet at the centre of
the Perseus cluster (B\"ohringer et al. 1993). Also note how (assuming a
`single galaxy' jet model) the northern jet appears to flare and brighten once
it reaches the lower density (\ie\ less bright) X-ray emitting gas. One might
have thought that the radio emission would expand and fade, on losing its
confining medium, i.e. once it entered a lower-density region. However it is
known (Loken \etal\ 1994) that a transition even from a dense to a less dense
medium can induce instabilities in radio jets, leading to shocks and
brightening. Though this may be the case, other possibilities to explain the
northern feature include the idea that the northern jet has perhaps run into
the northern galaxy, or that, as discussed earlier, there may be significant
radio emission from the northern galaxy itself - the `two galaxies' model.

\subsection{ATCA\,J0056.0-3723}

Lastly, we move on to the other bipolar twin-jet radio galaxy (\AT; Fig.~1),
lying 3$\arcmin$ east of \ESOs. We estimate the optical apparent magnitude of
the faint coincident source seen in the DSS2 image to be m$>$22, this based on
comparison with the nearby optical quasar QSO\,J005555.6-372219 with m=20.1
(Veron-Cetty \& Veron 1998) (a source is also listed in the SuperCosmos Sky
Survey Catalogue within 1$\arcsec$ of ATCA\,J0056-3723 with a magnitude B(J)
of 22.65). This optical faintness indicates that the source is likely a
distant unrelated background radio galaxy (radio galaxies are giant
ellipticals with a small range in absolute magnitude). Finally, it should be
stated that, while no X-ray emission was detected from this source, the source
is positioned close to the PSPC window support structure, which reduces
significantly the X-ray detection efficiency.

\section{Summary}

The ATCA observations of the peculiar galaxy pair \ESO\ at the centre of the
poor cluster \AB\ show what appears to be a bipolar, fairly assymetrical,
twin-jet radio galaxy, associated with the southern system \ESOs, itself a
galaxy merger. We also discuss the fact that an extra contribution from a
one-sided radio jet from the northern galaxy, \ESOn, may exist.

We estimate the jets, with projected lengths of up to 100\,kpc, 
to have velocities of at least 1000\,km s$^{-1}$,
and can explain the bending of the jets as being due to \ESOs\ moving through
the ICM at $\sim190$\,km s$^{-1}$. ROSAT PSPC observations
indicate relatively cool diffuse X-ray emission consistent with group or poor
cluster emission, and this emission, when compared with the radio jets, is
suggestive of channeling effects taking place, \ie\ the jets are able to
punch holes, and displace the X-ray emitting gas, as is seen in other systems.
Another bipolar jet radio galaxy discovered close by is likely to be a
background object.

\begin{acknowledgements}
We used the Karma/MIRIAD software package developed by the ATNF and the 
EXSAS/MIDAS software package developed by the MPE. We thank T. Panutti 
for considerable support during the radio continuum observations, and 
are very grateful to the referee for comments which greatly 
improved the paper. The ROSAT 
project is supported by the German Bundesministerium f\"ur Bildung und 
Forschung (BMBF) and the Max-Planck-Gesellschaft (MPG). 
Based on photographic data obtained using The UK Schmidt Telescope.
The UK Schmidt Telescope was operated by the Royal Observatory
Edinburgh, with funding from the UK Science and Engineering Research
Council, until 1988 June, and thereafter by the Anglo-Australian
Observatory.  Original plate material is copyright (c) the Royal
Observatory Edinburgh and the Anglo-Australian Observatory.  The
plates were processed into the present compressed digital form with
their permission.  The Digitized Sky Survey was produced at the Space
Telescope Science Institute under US Government grant NAG W-2166.

\end{acknowledgements}

\end{document}